\documentclass{article}

\usepackage{PRIMEarxiv}

\usepackage[utf8]{inputenc} 
\usepackage[T1]{fontenc}    
\usepackage{hyperref}       
\usepackage{url}            
\usepackage{booktabs}       
\usepackage{amsfonts}       
\usepackage{nicefrac}       
\usepackage{microtype}      
\usepackage{lipsum}
\usepackage{enumitem}
\usepackage{subfig}
\usepackage{amsmath}
\usepackage{fancyhdr}       
\usepackage{graphicx}       
\graphicspath{{media/}}     

\title{A tabletop Optically Pumped Magnetometer setup for
the monitoring of magnetic nanoparticle clustering and
immobilization using Thermal Noise Magnetometry
}

\begin{document}
\maketitle
\begin{center}
\vskip0.5\baselineskip{\bf Katrijn Everaert$^{1,2}$, Tilmann Sander$^{1}$, Rainer Körber${^1}$, Norbert Loewa${^1}$, Bartel Van Waeyenberge${^2}$, Jonathan Leliaert${^2}$, Frank Wiekhorst${^1}$ }
\small{\vskip0.5\baselineskip{\em$^{1}$Physikalisch-Technische Bundesanstalt – Department of Biosignals, Abbestrasse 2-12, 10587 Berlin, Germany\\
$^{2}$Ghent University – Department of Solid State Sciences, Krijgslaan 281, 9000 Ghent, Belgium }}
\end{center}

\begin{abstract}
Many characterization techniques for magnetic nanoparticles depend on the usage of external fields. This is not the case in Thermal Noise Magnetometry (TNM), where thermal fluctuations in the magnetic signal of magnetic nanoparticle ensembles are measured without any external excitation. This can provide valuable information about the fundamental dynamical properties of the particles, due to the purely observative experiments of this relatively new technique. Until now, TNM signals have been detected only by a superconducting quantum interference device (SQUID) sensor. We present a tabletop setup using Optically Pumped Magnetometers (OPMs) in a small magnetic shield, offering a flexible and accessible alternative and show the agreement between both measurement systems for two different commercially available nanoparticle samples. Due to the phase insensitivity of TNM, we are able to increase the OPM bandwidth of 100 Hz to 550 Hz by accounting for the frequency response profile of the magnetometers. We argue that the OPM setup with high accessibility complements the SQUID setup with high sensitivity and bandwidth. Furthermore, because of its excellent sensitivity in the lower frequencies, the OPM tabletop setup is well suited to monitor aggregation processes where the magnetization dynamics of the particles tend to slow down, e.g. in biological processes. As a proof of concept, we show for three different immobilization and clustering processes the changes in the noise spectrum measured in the tabletop setup: 1) the aggregation of particles due to the addition of ethanol, 2) the formation of polymer structures in the sample due to UV exposure, and 3) the cellular uptake of the particles by THP-1 cells. From our results we conclude that the tabletop setup offers a flexible and widely adoptable sensor measurement unit to monitor the immobilization and clustering of magnetic nanoparticles over time for different applications.
\end{abstract}


\section{Introduction}
The characterization of magnetic nanoparticles (MNPs) is a crucial part in the process towards their safe and efficient usage in biomedical applications\cite{Pankhurst2003,Tong2019,CoeneLeliaert2022}. Not only single particle properties such as size, shape and composition influence their magnetic behaviour, but also the state of the particles with respect to their environment. Clustering, aggregation, and immobilization of MNPs are processes of special interest in the context of biomedical applications\cite{Etheridge2014}. During arrival at the targeted body tissue, local particle concentrations get high and interparticle distances small. Therefore, magnetic exchange and dipolar interactions between particles may become important. Their mobility also changes during cellular uptake or molecular binding. These processes affect their magnetic properties\cite{DEberbeckFWiekhorst2006, Gutierrez2019} and - as a consequence - their diagnostic\cite{Loewa2013, Paysen2020} and therapeutic\cite{Bender2018b,Cabrera2018,Ortega-Julia2022} effect. To optimize the biomedical performance of the particles, it is thus necessary to determine the behavior of the particles within the body\cite{CoeneLeliaert2022} and map the clustering, aggregation and immobilization of the particles in their biomedical environment.

Magnetic properties and magnetization dynamics of MNP are often determined by measuring their response to an external magnetic field excitation\cite{Wiekhorst2012,Ludwig2013,Bui2022}. This external perturbation can potentially change the magnetic state of the sample, thereby influencing the outcome of the method. However, the magnetic moments of the MNP fluctuate at non-zero temperatures, and probing the corresponding induced magnetic noise allows one to obtain similar information about the inherent properties of the sample. The analysis of the fluctuation dynamics of MNPs is the idea behind a recently developed characterization technique \cite{Leliaert2015}, known as Thermal Noise Magnetometry (TNM). It is a unique method, since the sample is characterized while in an equilibrium state.

Compared to other characterization techniques, the signals in TNM are rather small (down to a few femtotesla) and thus require sensitive magnetic field sensors. Superconducting Quantum Interference Devices (SQUIDs), which were used in previous TNM studies\cite{Leliaert2015,Leliaert2017,Everaert2021}, have a well established reputation in magnetometry and biomedical applications. Their success is attributed to their excellent sensitivity, broad bandwidth, and durability\cite{Cohen1972,Burghoff2009,Korber2016}. However, they have the disadvantage that they require a liquid He infrastructure and a rather large sample-probe distance in the centimeter range caused by the mandatory thermal insulation between sensor and sample at ambient temperature. To increase the accessibility and consequently the adoption of TNM  thus broadening its application field, there is a necessity for more flexible measurement setups. For this purpose, Optically Pumped Magnetometers (OPMs) form an attractive sensor system\cite{Budker2007}. In this work, we present a tabletop TNM setup based on commercially available OPMs operating in a laboratory magnetically shield. The operational bandwidth of the used OPMs is limited by a phase shift above 100 Hz. However, the spectral measure used in TNM is phase insensitive. By accounting for the frequency response profile of the magnetometers, qualitatively correct measurements are ensured and the bandwidth of the sensors is increased to 550 Hz. We compare the TNM results of two MNP systems measured with the OPM setup with those obtained with an in-house developed SQUID-based system.

Employing OPM sensors offer an ideal measurement approach to monitor clustering and aggregation of MNPs by TNM over time. TNM is particularly sensitive to particle clustering because the amplitude of the signal increases with the square of the volume of the individual fluctuators\cite{Everaert2021}, which means that an aggregate of two particles results in a signal that is twice as large as the sum of their individual signals. Moreover, no external excitation is required during the TNM measurement procedure, which could influence the clustering process itself or falsely influence the outcome of the measurement. Second, the excellent low-frequency performance of OPMs favors particle systems with slow dynamics or processes which tend to slow down the dynamics of the magnetic entities in the sample, such as clustering and immobilization processes. Finally, the setup can be operated anywhere with a conventional wall socket AC power due to the OPM flexibility and can thus also be used to track processes that require environmental and experimental freedom. As a proof-of-concept, we designed and performed three different experiments in the tabletop setup, which concern the clustering, aggregation, and immobilization of Perimag particles.

\section{Methods}
\subsection{TNM model}
Two mechanisms are responsible for the thermal fluctuations measured in TNM. In liquid samples, the particles are submissive to Brownian motion. The MNPs, and thus their magnetic moments, rotate at time scales\cite{Debye1929}
\begin{equation}
  \quad
    \tau_{\mathrm{B}} = \frac{3\eta V_{h}}{k_{B}T}
    \label{eq:brown}
\end{equation}
with $\eta$ the viscosity of the fluid, $V_h$ the hydrodynamic volume of the particle, and $k_BT$ the thermal energy in the system. Additionally, magnetization can also change within the frame of the particle itself, which is the only mechanism present if the particles are immobilized. The N\'{e}el fluctuation time depends Arrhenius-wise on the energy barrier $KV_c$ set by the anisotropy of the particle
\begin{equation}
  \tau_{\mathrm{N}} = \tau_0 \exp\left(\frac{KV_c}{k_{B}T}\right)
  \label{eq:neel}
\end{equation}
where $K$ is the anisotropy constant, $V_c$ the magnetic core volume, and $\tau_0$ the characteristic attempt time\cite{Brown1963}. The effective fluctuation time then naturally combines to
\begin{equation}
\tau_{\mathrm{eff}}=\frac{\tau_{\mathrm{N}} \tau_{\mathrm{B}}}{\tau_{\mathrm{N}}+\tau_{\mathrm{B}}},
\end{equation}
in samples where both mechanisms are present.  Depending on the size of the particles, $\tau_N$ or $\tau_B$ are dominant and define the value of $\tau_{\mathrm{eff}}$.\\

The magnetic time signal $B$ measured in TNM is stochastic in nature with an autocorrelation function
\begin{equation}
    G_B(t)=\langle B(0)B(t)\rangle = \langle B^2\rangle \exp(-\vert t\vert/\tau_{\mathrm{eff}}).
    \label{eq: autocorrelation}
\end{equation}
The Power Spectral Density (PSD) is then obtained from the Wiener-Khintchine theorem as the Fourier transform of the autocorrelation function\cite{Wiener1930,Khintchine1934}:
\begin{equation}
    S_B(f)=2\left(\frac{\mu_0 M V_c}{4\pi d^3}\right)^2 \cdot \frac{   (4\tau_{\mathrm{eff}})^{-1}   }{(\pi f)^2+(2\tau_{\mathrm{eff}})^{-2}}
    \label{eq: lorentzian}
\end{equation}

The amplitude of the fluctuations depends on the total magnetic moment of the sample $M V_c$, and the distance $d$ from the sample at which the magnetic field is measured. At typical distances of few mm to few cm, the TNM signal a of nanoparticle ensemble ranges from pT-fT.

\begin{figure}[h!]
  \centering
    \includegraphics[width=1\textwidth]{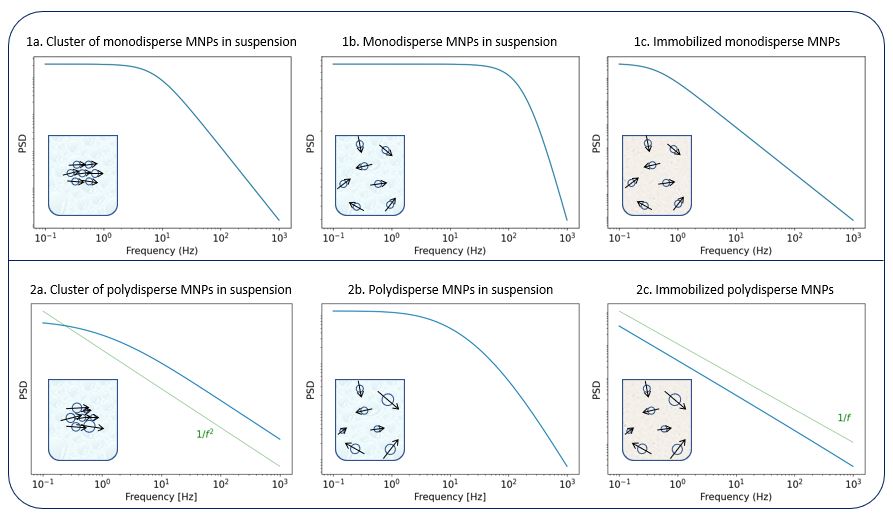}
    \caption{Theoretical noise spectra of different samples. The monodisperse PSDs are displayed for particles with a hydrodynamic diameter of $d_h=130$ nm and a core diameter of $d_c$ = 25 nm. the PSD is flat up to the characteristic cutoff frequency, after which it falls off with $1/f^2$. The diameters of the polydisperse particles follow a lognormal size distribution with parameters $d_h\sim$ ln N($\mu$ = 124 nm, $\sigma$=0.35) and $d_c\sim$ ln N($\mu =$25 nm, $\sigma$=0.45). In the case of polydisperse cluster formation (2a), the PSD still prominently has a $1/f^2$ shape. In the case of the immobilization of the polydisperse particles however, a typical $1/f$ shape is distinguished due to the extreme broad fluctuation time distribution. The clustered were taken to have 3 times the size of the single sinlge particles.}
    \label{Fig:Cartoon}
\end{figure}

The dynamics of the particles is quantified in the fluctuation time $\tau_\mathrm{eff}$, which therefore also defines the width of the PSD. The cutoff frequency $\nu_\mathrm{cutoff}=\frac{1}{2\cdot \tau_\mathrm{eff}}$ divides the PSD into two regimes: a low frequency regime where $f < \nu_\mathrm{cutoff}$ and a high frequency regime where $f > \nu_\mathrm{cutoff}$. The low frequency regime corresponds to the region on Fig. \ref{Fig:Cartoon}.1b. where the PSD is flat. In the high frequency regime, the PSD drops with $1/f^2$.

Direct parameters influencing the cutoff frequency are those entering equations (\ref{eq:brown}) and (\ref{eq:neel}), such as the suspension viscosity, the anisotropy constant, and the local temperature. However, a change in aggregation state or mobility of the particles, or an increase in the interparticle interaction, affects their magnetization dynamics - and thus the noise spectrum - as well. Fig. \ref{Fig:Cartoon} shows the theoretical expression of the PSD for different MNP configurations to illustrate changes in the noise spectrum. Particle clustering (Fig \ref{Fig:Cartoon}.1a) increases the hydrodynamic volume of the individual fluctuators with a decrease in the Brownian fluctuation time as a result. The cutoff frequency shifts towards smaller frequencies, and the $1/f^2$ behavior becomes more pronounced. The N\'{e}el mechanism is the submissive mechanism in most MNP systems (for the common case of iron oxide MNPs with rather large core diameters $d_c$>10 nm) . i.e. the N\'{e}el fluctuations are often orders of magnitude slower than Brownian rotations. This means that the cutoff frequency also shifts towards lower values upon elimination of Brownian rotations during immobilization (Fig. \ref{Fig:Cartoon}.1c).

For a polydisperse sample, the volumes $V_h$ and $V_c$ - and therefore also the fluctuation time $\tau_\mathrm{eff}$ - are distributed along $P(\tau_\mathrm{eff})$. The PSD can then be written as a superposition of independent fluctuators (\ref{eq: lorentzian})
\begin{align}
    S_b^\mathrm{poly}(f)&=\int_0^{\infty} P(\tau_\mathrm{eff})\cdot S_b^{\tau_\mathrm{eff}}(f)d\tau_\mathrm{eff}\\
    &=\int_0^{\infty}\int_0^{\infty}P(V_h)P(V_c)\cdot S_b^{V_h,V_c}(f)dV_{h}dV_{c} \label{Eq:size_dist}
\end{align}
This typically results in a stretching of the cutoff frequency $\nu_\mathrm{cutoff}$ over a certain frequency range, and a less distinct shape of the PSD as visible in Fig \ref{Fig:Cartoon}.2b. For a fairly broad size- or fluctuation time distribution, the PSD finally can be approximated by a 1/f curve\cite{Weissman1988, Leliaert2015}.

Similar to clustering of monodisperse particles, the hydrodynamic size of a polydisperse cluster also increases, and its related cutoff frequency shifts towards lower frequency values. The $1/f^2$ falloff dominates in the considered bandwidth as shown on Fig. \ref{Fig:Cartoon}.2a. The effect of immobilization of a polydisperse sample is even more pronounced, since the N\'{e}el fluctuation time depends exponentially on the core volume. The size distribution is stretched into a broad fluctuation time distribution, and the PSD gets the distinct $1/f$ shape which is displayed on Fig. \ref{Fig:Cartoon}.2c.

We would like to point out that the clustering, aggregation, and immobilization of MNPs are generally not uncorrelated and often occur at the same time. The broad size distributions $P(V_h)$ and $P(V_c)$ also make the quantitative interpretation of the fluctuation dynamics of the magnetic moments less straightforward than for the model curves displayed in Fig. \ref{Fig:Cartoon}.

\subsection{Experimental setups}
\subsubsection{Magnetic Nanoparticles.}
Two commercially available MNP systems have been used for the comparison of the two setups: Resovist particles (an MRI liver contrast agent provided by Meito Sanyo, Japan) with an iron concentration of $c$(Fe)=429.1 mmol/L and Perimag particles (Micromod Partikeltechnologie GmbH, Rostock, Germany) with a plain surface and an iron concentration of $c$(Fe)=644.4 mmol/L.  Perimag particles were chosen for the proof-of-concept experiments with a COOH group coating for cellular uptake.

\subsubsection{OPM tabletop setup for Thermal Noise Magnetometry.}
Altough the concept of magnetic sensing by use of optical pumping dates back to 1950-1960, the field of Optically Pumped Magnetometers is still developing steadily. In this technique, an alkali metal gas vapor - often Rb or Cs - is polarized by pumping with a polarized light beam. Once fully polarized, the gas becomes transparent. A magnetic field changes the polarization state of the vapor atoms, which is quantified by measuring the polarity or intensity of a second probing light beam through the gas vapor. Today, there are many different OPM configurations, covering a broad range of applications\cite{Sander2020, Bason2022, Deans2021}. Comparisons have been made with SQUID systems has been made\cite{Knappe2010, Marhl2022} and the magnetometers have also found their way into applications such as MNP characterization and imaging\cite{Johnson2012,Dolgovskiy2015,Baffa2019,Jaufenthaler2020a,Jaufenthaler2021}.

The developed tabletop setup consists of two Gen-2 QuSpin Zero-Field Magnetometers (QZFMs) (QuSpin Inc., CO, USA)\cite{Shah2013} that are operated in single-axis mode. One is placed near the MNP probe (QZFM1), the other is used for reference measurements (QZFM2).

The sample and the QZFMs are placed inside a laboratory MS-2 magnetic shield (Twinleaf LLC, NJ, USA) to minimize the effect of external fields on the MNPs dynamics and to ensure the proper working of the QZFMs. The shield consists of four metal layers and has a shielding factor of $10^6$ as specified by the manufacturer\cite{twinleaf}. The controlling QZFM electronics is placed outside the shield and driven via the program QuSpin ZFM on a laptop, from which the data is also collected with a U6 Labjack (LabJack Corporation, CO, USA) in stream mode.
\begin{figure}[h!]
  \centering
    \includegraphics[width=1\textwidth]{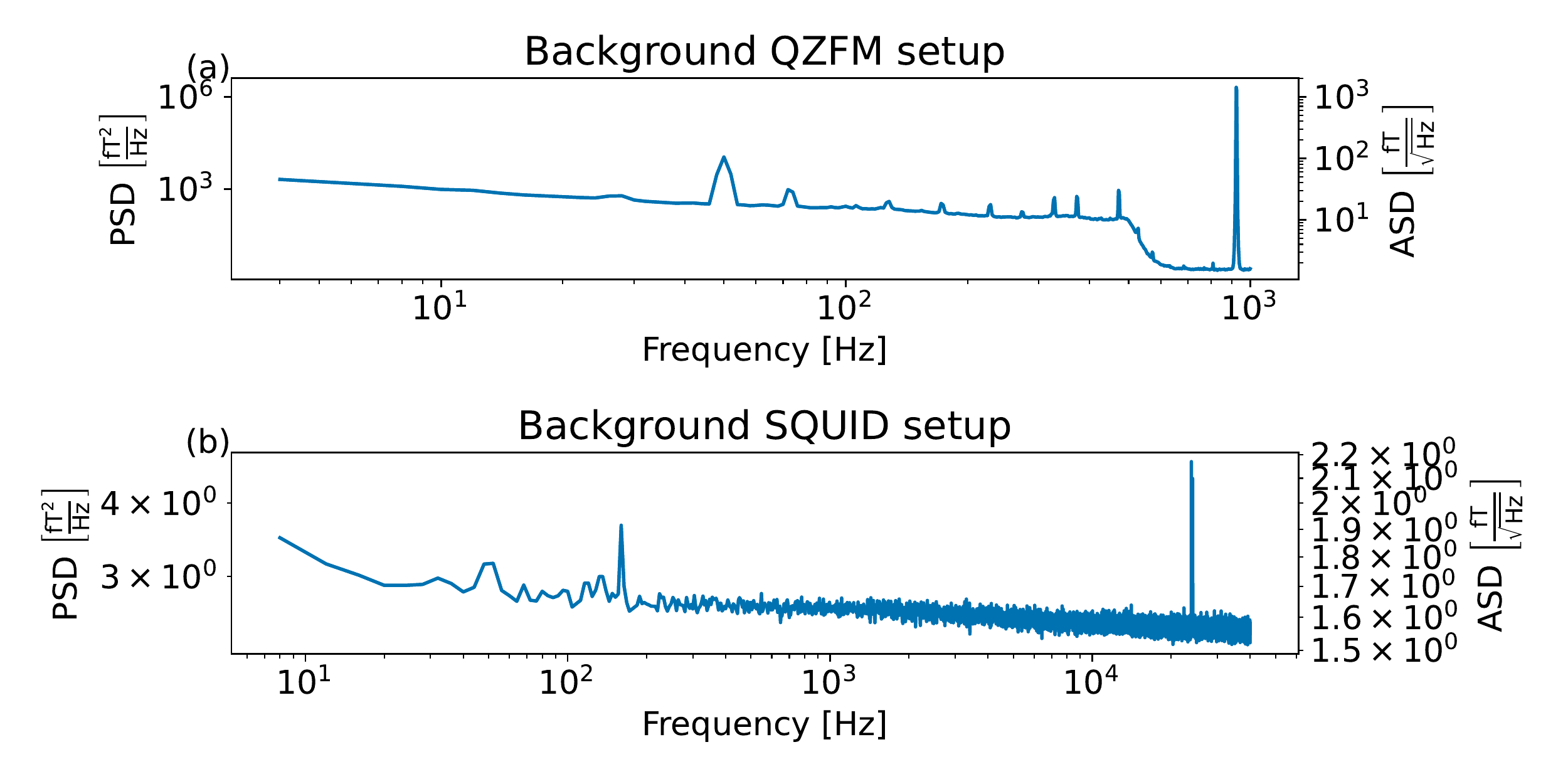}
    \caption{Background power spectral density in tabletop QZFM setup (a). The 50 Hz contamination from the power line and its related 150 Hz peak are visible amongst other environmental disturbances. The 923 Hz signal from the QZFM modulation and its aliasing peak around 77 Hz . (b) Background of the in-house developed SQUID setup. The 50 Hz and 150 Hz peaks from power line residuals are visible, as well as a two peaks around 24 kHz. }
    \label{Fig:BG_both}
\end{figure}
The tabletop setup was operated in a conventional lab environment at the Physikalisch-Technische Bundesanstalt in Berlin. The 50 Hz contamination from the power line and its related 150 Hz peak are visible in the measured background spectrum on a noise floor of 200-2000 $\frac{\textrm{fT}^2}{\textrm{Hz}}$ (approx. 10-30$\frac{\textrm{fT}}{\sqrt{\textrm{Hz}}}$). Among other environmental disturbances, the 923 Hz signal from the QZFM modulation and its aliasing peak around 77 Hz are visible\footnote[1]{These disturbances can be minimized further by placing the shield in an aluminum cage, and grounding both the shield and the cage to the USB ground of the laptop.}. The manufacturer does not specify any product information of the QZFM above 100 Hz, because of the phase shift in the signal above this frequency. However, since our spectral measure is phase-insensitive, we are able to extend the bandwidth of the QZFM beyond their usual 100 Hz frequency range, as explained in Sec. \ref{Sec:freq_resp}. Time signals of up to 20 minutes were recorded at a sample rate of 2 kHz, and the PSD were calculated and averaged as explained in the method section of Ref. \cite{Everaert2021}. The final displayed spectra are calculated by subtracting the background spectrum PSD$_\mathrm{BG}$ from the MNP spectrum PSD$_\mathrm{MNP}$, i.e. the spectra of an MNP sample that were measured in the tabletop setup.

To achieve an optimal Rb density for increased sensitivity, the QZFM vapor cell is heated to a temperature of about $160 	^{\circ} $C\cite{Shah2013,Borna2017}. This has an immediate effect on the temperature of the sample, and overheating of the MNPs is prevented by placing a 2 mm thick isolation material between the housing of the QZFM and the sample. A sample temperature of $43.2 \pm 0.5 	^{\circ} $C was measured after a stabilization time of 20 min after placement in the setup. With the isolation material included, the minimal distance between the centre of the vapor cell and the sample is estimated at 8.5 mm. With a sample height of approximately 10 mm, the average distance between the centre of the vapor cell and the particles is 13.5 mm.

\subsubsection{SQUID setup for Thermal Noise Magnetometry.}
The in house developed SQUID setup consists of a superconducting Niobium shield\cite{Ackermann2007} in which 6 SQUID sensors with a rectangular pickup coil are operated. The sample can be placed inside a warm bore at an average distance of 23.5 mm from the pickup coils. Only one sensor is used for the TNM experiment.
The background spectrum of the SQUID setup shows a relatively flat profile with values between 2-3.5 $\frac{fT^2}{Hz}$ (1.5-1.8 $\frac{fT}{\sqrt{Hz}}$). 50 Hz and 150 Hz peaks from power line residuals are visible, as well as a two peaks around 24 kHz.
\begin{figure}
    \centering
    \subfloat[\centering ]{{\includegraphics[width=0.45\textwidth]{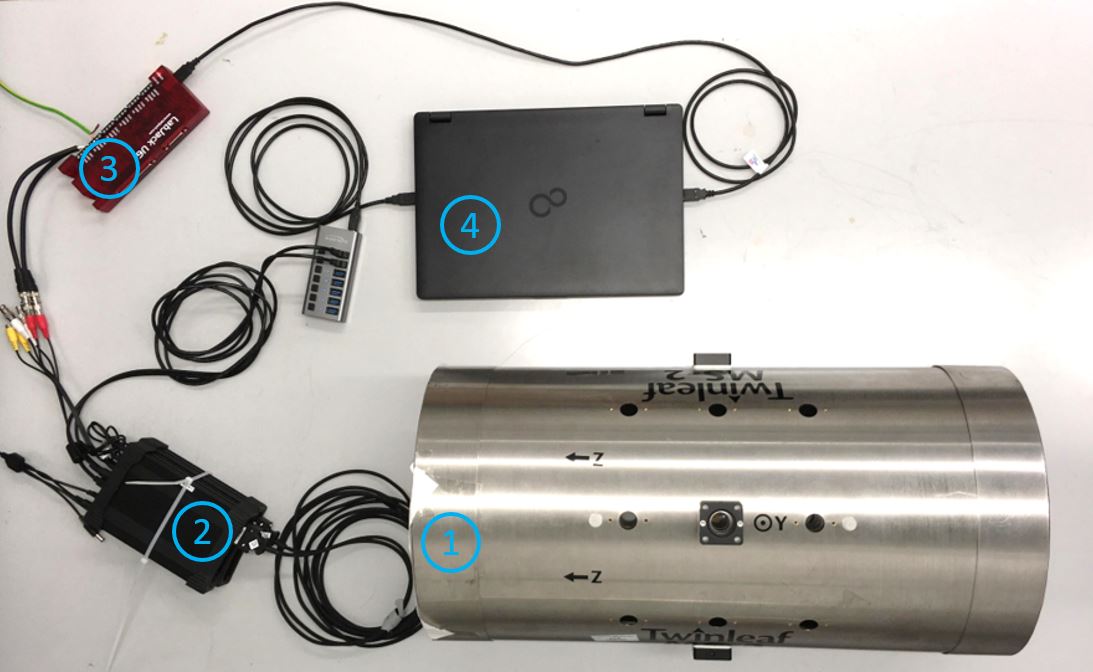} }}%
    \qquad
    \subfloat[\centering ]{{\includegraphics[width=0.45\textwidth]{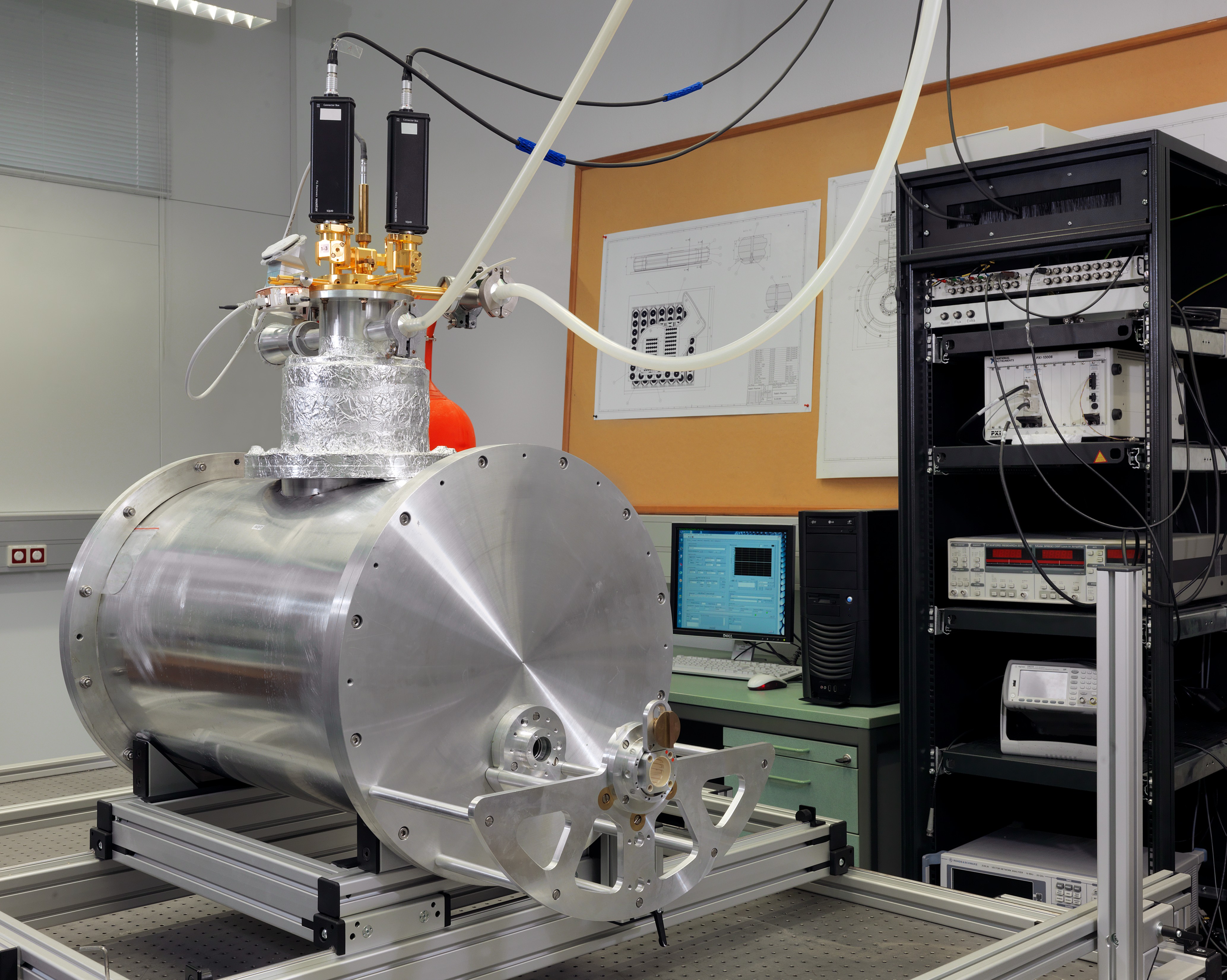} }}%
    \caption{(a) Tabletop TNM setup based on OPMs. The sample is places inside the Twinleaf MS-2 shield (1) together with the two QZFMs. QZFM electronics (2) controle the sensors and a laptop (4) serves for driving the DAQ Labjack U6 (3) in stream mode, running the QZFM electronics and collecting the data. The sample is on an average disntance of 13.5 mm from the sensible volume.\\(b) In-house developed SQUID setup for reference measurements. A superconducting Nb magnetic shield 6 SQUID sesors are kept on LHe temperature, the sample is placed in a warmbore and kept on an average temperature of $43.0 \pm 0.5 ^{\circ} C$ to match the sample temperature in the tabletop setup. An average distance of 23.5 mm is measured between the pickup coils of the SQUID sensors and the sample. Only one sensor is used for the TNM measurement.}
    \label{fig:setup}
\end{figure}
As is clear from Eq. (\ref{eq:brown}) and Eq. (\ref{eq:neel}), the dynamics of the particles is strongly dependent on the temperature. To compare both measurement systems under equal conditions, the sample has been kept at a constant temperature of $43.0 \pm 0.5 ^{\circ} C$ in the SQUID setup to match the sample temperature in the QZFM setup. This was achieved by use of a stable airflow through the warm bore. 13 minute time signals were acquired at a sample rate of 100 kHz and the PSDs were subsequently calculated and averaged.

\subsection{Frequency response of the sensors}
\label{Sec:freq_resp}
\begin{figure}[h!]
  \centering
    \includegraphics[width=1\textwidth]{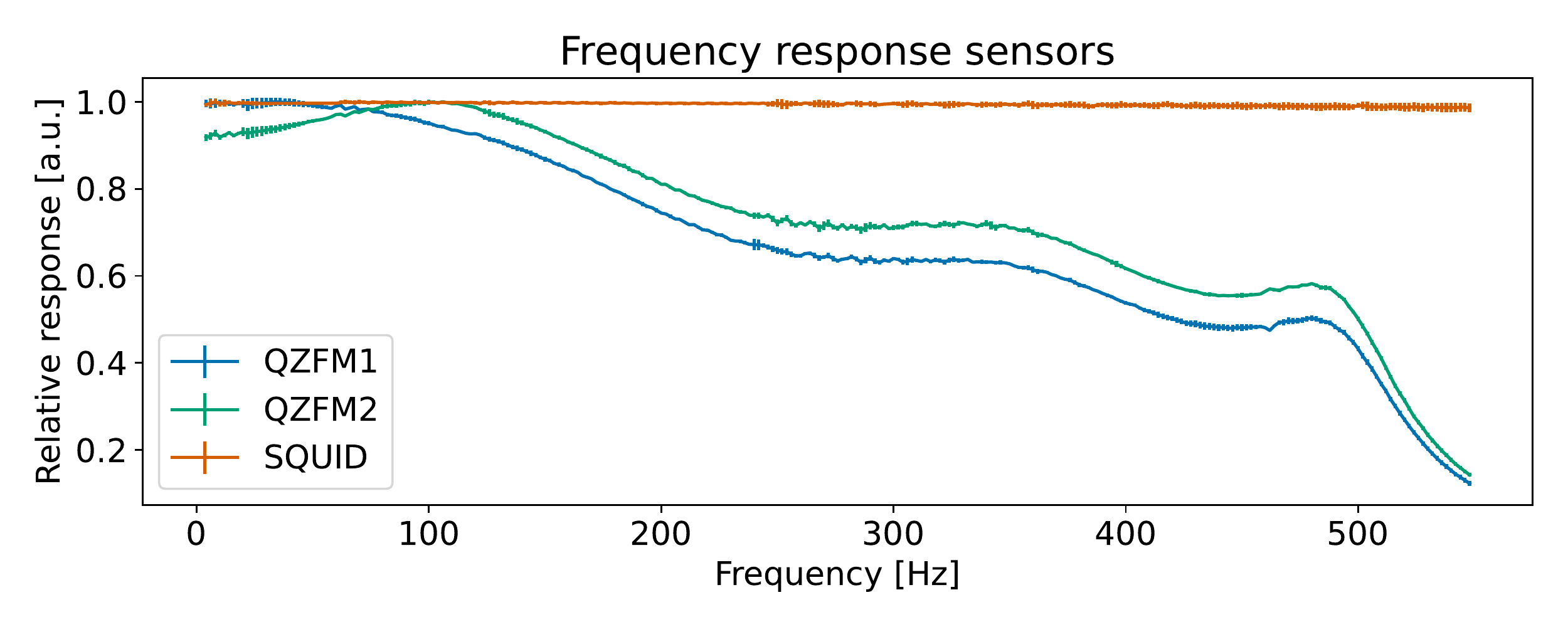}
    \caption{Measured frequency responses $H(f)$ of the different sensors. The data was acquired by application of an AC magnetic field with sweeping frequency.}
    \label{Fig:freq_resp}
\end{figure}
To increase the QZFM frequency range above 100 Hz, a frequency response profile of the QZFMs has been measured to ensure a quantitatively correct measurement of the power spectra of the MNPs. To this end, the sensors were placed inside a magnetic shielded room at the Physikalisch-Technische Bundesanstalt (named as the “Zuse-MSR” in Ref. \cite{Voigt2015}) and a homogeneous AC field with an amplitude of 772 pT was applied by use of a square Helmholtz coil. Its frequency was swept in the range of $[2-600]$ Hz. The amplitude of the QZFM output signal was monitored and analysed in the time domain, after which the response values were averaged and normalized. From this data set, a frequency response profile was calculated for both QZFMs as shown in Fig. \ref{Fig:freq_resp}. The uncertainty on the response values was calculated from the standard deviation of the different peaks of the excitation and the response at one frequency. Note that the frequency response varies strongly and differently for both QZFMs. It is also not constant within the low frequency measurement range of OPM applications. For comparison, a similar procedure was performed with the SQUID sensor with a field of 751 nT by inserting a small coil into the warm bore. There was no frequency dependence of the amplitude detected as can be seen in Fig. \ref{Fig:freq_resp}.

Having determined the frequency response of the QZFMs, the Power Spectral Density $S_B(f)$ of a MNP ensemble was calculated as
\begin{equation}
    S_B(f)=\frac{S_{QZFM}(f)}{H(f)^2}
\end{equation}
with $S_{QZFM}(f)$ the Power Spectral Densities measured by the QZFMs and $H(f)$ their relative frequency responses as displayed in Fig. \ref{Fig:freq_resp}.

\section{Results and discussion}
\subsection{Comparison of Power Spectra}
\begin{figure*}[h!]
  \centering
    \includegraphics[width=1\textwidth]{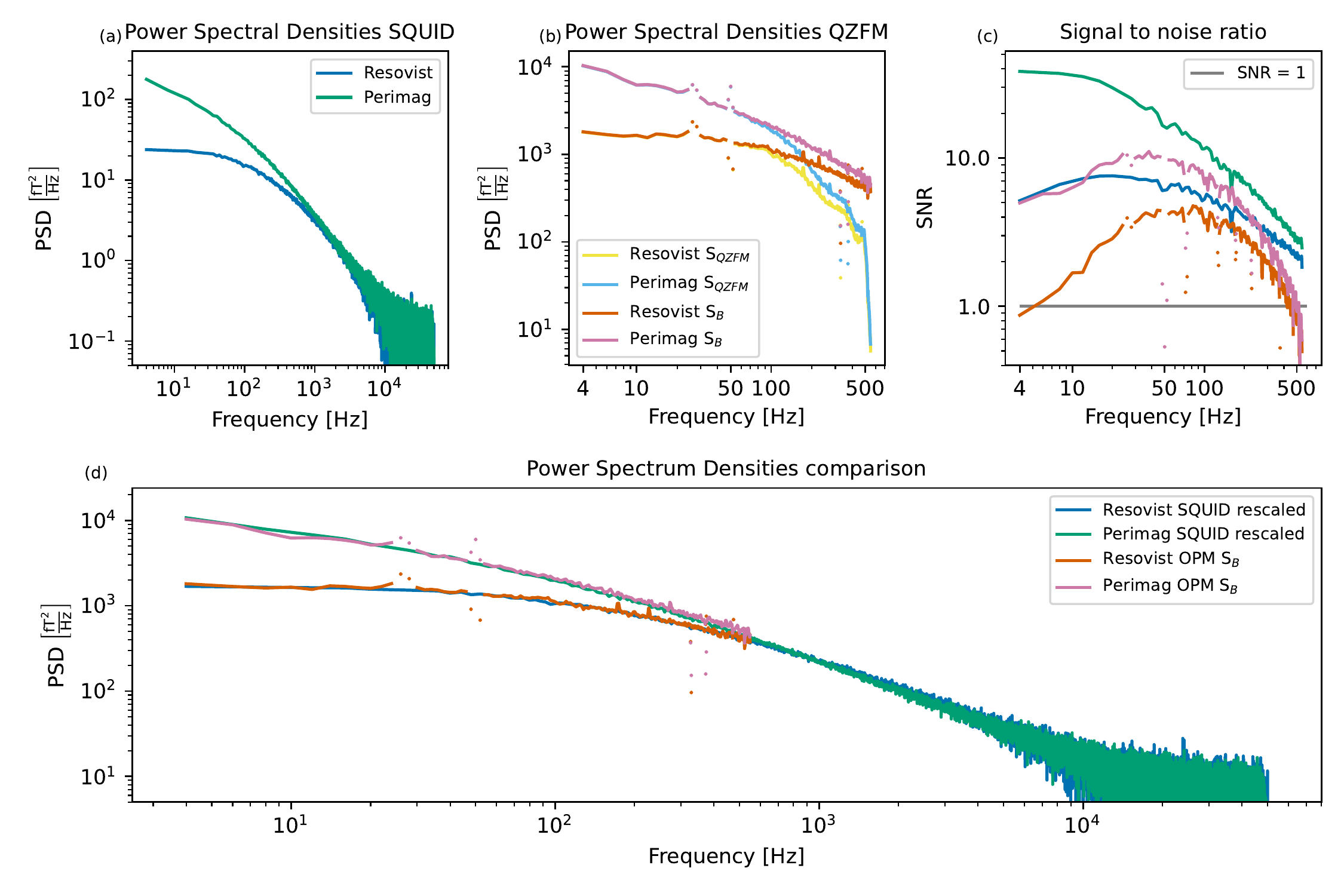}
    \caption{Measured SQUID profiles of the two MNP systems (a) and measured and compensated QZFM profiles of the two MNP system (b). For clarity, the points in the QZFM spectra at the unstable background frequencies (30, 50, 330, 375 Hz) have been plotted separately. These corresponds to the peaks in the background spectrum of the tabletop setup in Fig. \ref{Fig:BG_both}. Qualitative comparison of the MNP spectra measured in both setups (c). The SQUID spectra of (a) have been rescaled to match the power of the OPM spectra S$_B$ in (b) at 80 Hz. Apart from the unstable background peaks in the OPM spectra, a very good agreement between the measurements obtained in both setups is visible.}
    \label{Fig:Comparison}
\end{figure*}

Fig. \ref{Fig:Comparison} (a) shows the Power Spectral Densities of both MNP systems measured in the SQUID setup. The spectra displayed are the noise spectra measured with MNPs PSD$_\mathrm{MNP}$ with the background spectra PSD$_\mathrm{BG}$ subtracted. The raw spectra can be found in the Supplementary Material. The PSD of the Resovist system is relatively flat up to a cutoff frequency of about 90 Hz, after which the PSD starts to decrease continuously due to the size distribution of the particles. On the other hand, the Perimag system shows higher power at lower frequencies, with cutoff frequency located at values lower than the displayed bandwidth. This is a result of the slower magnetization dynamics due to the large hydrodynamic size of the Perimag particles with a broad size distribution, as explained by Eq. (\ref{Eq:size_dist}). If the PSDs are normalized to iron the amount in the samples (see the Supplementary Material), a crossing occurs around 200 Hz.

The same MNP systems are measured in the tabletop setup, where a bandwidth up to 550 Hz can be reached. The spectra before ($S_{QZFM}(f)$) and after ($S_B(f)$) the frequency response correction are displayed in Fig. \ref{Fig:Comparison} (b).

Due to the reduced sample-sensor distance, the MNP signal is higher in the tabletop setup than in the SQUID system. However, the tabletop setup is less sensitive than the SQUID setup. This is clear from Fig. \ref{Fig:Comparison} (c), which shows the signal-to-noise ratio (SNR) as a function of frequency for both setups and both MNP systems:
\begin{equation}
SNR(f)=\frac{PSD_{MNP}(f)-PSD_{BG}(f)}{PSD_{BG}(f)}
\label{Eq:SNR}
\end{equation}
The tabletop setup has a steeper SNR loss than the SQUID setup above 100 Hz. Not only the signal, but also the noise is amplified as a result of the frequency response compensation. The excellent state of the SQUID setup becomes visible in the SNR plots. Both MNP systems have a SNR up to one order of magnitude higher in the SQUID setup compared to the tabletop OPM setup, even if the signal is two orders of magnitude lower.

SNR = 1 marks the limit where environmental and sensor noise (that is, unwanted noise) has the same amplitude as MNP noise (that is, wanted noise). Although this is not a strict limit of acceptance, it can still serve as a mark to validate the SNR of the MNP systems and the setup. Only Perimag will be used for the proof-of-concept experiments, since the SNR of Resovist is relatively low in the lower frequency range in the tabletop setup. Moreover, the SNR of Perimag above 400 Hz also crosses the SNR=1 limit.  For lower concentrated samples, such as those used in the proof-of-concept experiments, this crossing will occur even at lower frequencies.

The measurements are directly compared in Fig. \ref{Fig:Comparison} (d) by rescaling the SQUID spectra to match the OPM spectra at 80 Hz. As the curves of the two particle systems measured in two setups overlap nicely, we conclude that the compensation for the frequency response profile of the QZFM is a valid approach. Despite their loss in sensitivity above 100 Hz, the QZFMs recover a quantitatively correct spectrum. For MNP samples with high power in the lower frequency range, such as the Perimag and Resovist samples used as example MNP systems here, both measurement systems are thus equally suitable. For smaller MNP systems with dynamics in the higher frequency range, the tabletop setup might not be sufficient, both in bandwidth and sensitivity. However, these particle systems could still be tuned to fall within the QZFM bandwidth by increasing the viscosity of the suspension, as proposed in Ref. \cite{Leliaert2017}.

To further increase the bandwidth, both setups still have some potential towards lower frequencies. Since both magnetic shields (i.e. the superconducting shield integrated into the dewar of the SQUID system and the mu-metal Twinleaf shield of the tabletop setup) are relatively small, the low-frequency shielding is very effective. Longer measurement times then lead to larger time intervals for the Fourier transformation and averaging procedure, reaching lower frequencies in the spectra. For an increase towards higher frequencies, the SQUID setup could measure at higher sample rates. However, the SNR also gets small at 50 kHz, with relatively little information gain for these MNP systems. The current tabletop setup is limited to 550 Hz due to the frequency of the modulation signal of the QZFMs.

\subsection{Monitoring of clustering processes}
Since the TNM signal scales quadratically with the volume of the noise sources \cite{Everaert2021}, this technique is particularly suited to monitor clustering processes of magnetic nanoparticles. The absence of any driving field during the measurement also excludes any undesired effects induced by an external excitation. Moreover, the good performance of the QZFM at lower frequencies favours processes which tend to slow down the magnetization dynamics of the sample. An OPM based TNM setup thus offers a broadly applicable tool to monitor the clustering of MNPs. As a proof of concept, we report on the monitoring of three such processes, measured with TNM in the described tabletop setup:
\begin{enumerate}
    \item Enforced aggregation of Perimag particles by addition of ethanol
    \item Formation of photopolymer structures in a Perimag sample by exposure to UV light
    \item Cellular uptake of Perimag particles by THP-1 cells 
\end{enumerate}

\subsubsection{Enforced aggregation of Perimag particles by addition of ethanol}
In a first example, the aggregation of Perimag particles is enforced by adding ethanol to the sample. A 200 $\mu$l Perimag plain solution with an iron concentration of c(Fe)=466.4 mmol/L was diluted with 200 $\mu$l ethanol. The stabilizing dextran surfaces of the particles are dissolved, the attractive forces between the magnetic cores prevail, and the system aggregates. The sedimentation of the aggregates due to gravity was visually detectable after several seconds.
\begin{figure}[h!]
  \centering
    \includegraphics[width=1\textwidth]{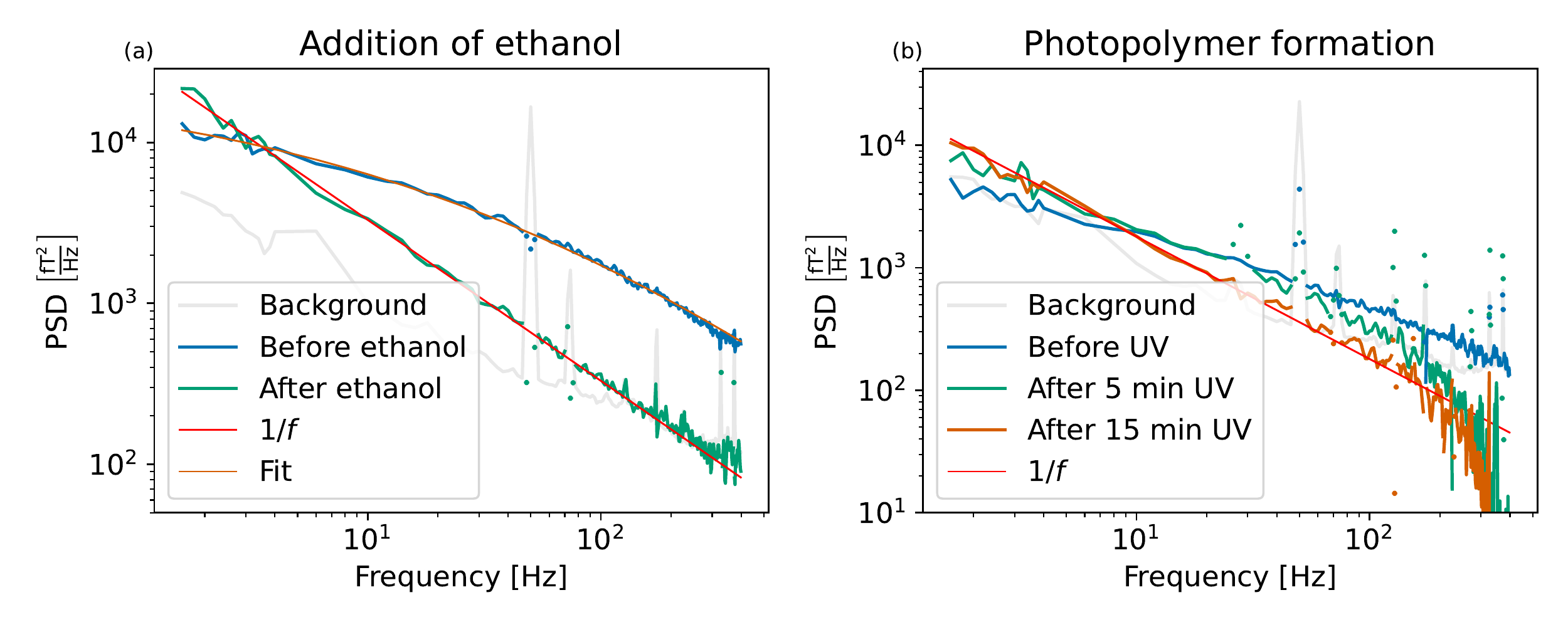}
    \caption{(Left) Power Spectral Densities of Perimag particles before and after addition of ethanol. A lognormal size distribution was fitted to the MNP sample before aggregation. After the addition of ethanol, the magnetic cores aggregate and sediment due to gravity. Due to the extremely broad distribution of the N\'{e}el fluctuation times, the PSD has the distinct $1/f$ shape as shown in Fig \ref{Fig:Cartoon}.2c. (Right) Power Spectral Densities of Perimag particles before and after polymer formation. A gradual immobilization of the particles is induced during UV curing and full immobilization is reached after 15 minutes exposure time.}
    \label{Fig:ethanol_polymer}
\end{figure}
The influence of aggregation on the noise spectrum of Perimag is visible in Fig. \ref{Fig:ethanol_polymer} (a), where a spectrum before and after aggregation is displayed. Since the geometry of the sample is not conserved due to a change in spatial distribution of magnetic material, the spectra show only qualitative effects. A lognormal size distribution logN($\mu$=72.6 $\pm$5 nm, $\sigma$=0.82 $\pm$ 0.3) with an average hydrodynamic diameter of 101 $\pm$ 26.0 nm was fitted to the curve before addition of ethanol. Given the limited bandwidth of 400 Hz, these parameters match the average diameter of 130 nm of the manufacturer reasonably well.

After the addition of ethanol, the aggregates sediment and are only submissive to the N\'{e}el mechanism. Their noise curve is dominated by $1/f$ noise. The slow magnetization dynamics of the aggregated cores and the broad size distribution of their fluctuation times are a distinctive signature of this process which was cartoonized in Fig \ref{Fig:Cartoon}.2c.

\subsubsection{Formation of photopolymer structures in a Perimag sample by exposure to UV light}
Photopolymer resins are popular materials used in additive manufacturing. They form a highly controllable system to gradually solidify suspensions. In combination with magnetic nanoparticles, they are of particular interest for precise phantom design and fabrication\cite{Loewa2021,Nordhoff2022} to evaluate Magnetic Particle Imaging scanners\cite{Loewa2019,Dutz2019}. In our experiment, a photopolymer was mixed with Perimag particles to mimic the gradual change in mobility of the particles when being embedded in the target tissue.

First, the Perimag-photopolymer mixture was prepared by adding 100 $\mu$L Perimag plain with an iron concentration of 644.4 mmol/L to a 100 $\mu$l photopolymer base material \footnote[2]{Perfactory acrylic R5 red from EnvisionTEC Inc., composed of acrylic acid esters and a photoinitiator (0.1 $-$ 5 \text{\%}). The Perfactory Acryl R5 resin has a density of 1.12 $-$ 1.13 g/cm$^3$.} in a 2 ml Eppendorf tube. A homogeneous spatial distribution of the particles in the base material was ensured by sonication with an ultrasound sonifier (UP200Ht, Hielscher Electronics, Germany). 120 $\mu$l of the mixture was used as a sample and a first spectrum was measured before UV exposure. The sample was then exposed to UV light in a UVACUBE 2000 for 5 and 10 min subsequently.

Fig. \ref{Fig:ethanol_polymer} (b) shows the measured spectra of the particles in the base material before exposure, and after 5 and 15 min total exposure time. Since only 60 $\mu$l magnetic material has been used in this experiment, the TNM signal is lower than in the spectra shown previously. Therefore, we argue that the falloff above 200 Hz of the two exposure spectra is an artificial effect due to insufficient SNR and not the physical shape of the spectra, which we expect to decrease linearly on the log-log scale.

Before UV exposure, the particles rotate freely in the highly viscous base material. Compared to the water suspended particles in Fig. \ref{Fig:ethanol_polymer} (a), their Brownian rotations are slower. The related cutoff frequency is shifted to lower frequencies outside the window, and only the straight tail of the PSD is visible. Brownian movement of the particles is further excluded due to crosslink formation as the MNPs get enclosed in small polymer cavities during UV curing. The effective viscosity increases towards an eventual full immobilization. The Brownian fluctuations gradually slow down, the related Brownian cutoff frequency moves closer towards DC values and N\'{e}el fluctuations become dominant. After 15 minutes of exposure time, the PSD reaches the limiting $1/f$ shape on Fig. \ref{Fig:ethanol_polymer} (b). All particles are immobilized, as the PSD is directly comparable with the PSD of the aggregates in Fig. \ref{Fig:ethanol_polymer} (a).

During UV curing, volume and geometry of the sample are conserved and the spectra can be compared quantitatively. This allows us to define an effective immobilization degree based on the PSD value at a stable low frequency after each exposure step. A normalization of the PSD values at 1.6 Hz to the fully immobilized state gives an immobilization of 50\% before UV exposure and 72\% after five minutes exposure time. This experiment therefore shows the potential of TNM to be used for continuous monitoring during MNP clustering and immobilization processes.
\subsubsection{Cellular uptake of Perimag particles by THP-1 cells}
MNPs are known to form clusters during cellular uptake, which impacts their magnetization dynamics\cite{Loewa2013,Etheridge2014,DiCorato2014,Poller2016,Teeman2019}. For their usage in biomedical application as Magnetic Particle Imaging (MPI) and hyperthermia treatment, the change in their magnetic state can heavily influence their performance\cite{Bender2018b,Cabrera2018,Paysen2020,Remmo2022}. However, the change in the thermal noise of the MNPs due to cellular uptake is unknown so far. Especially the absence of an external magnetic excitation during a TNM experiment can be seen as an advantage in the determination of the precise clustering mechanism, since cluster formation and aggregation due to an external perturbation are eliminated in this technique. In a third experiment, the noise profile of COOH coated Perimag particles is measured after cellular uptake by THP-1 cells in the tabletop setup and compared with the pre-uptake water suspended system.

\begin{figure}
    \centering
    \subfloat[\centering ]{{\includegraphics[width=0.45\textwidth]{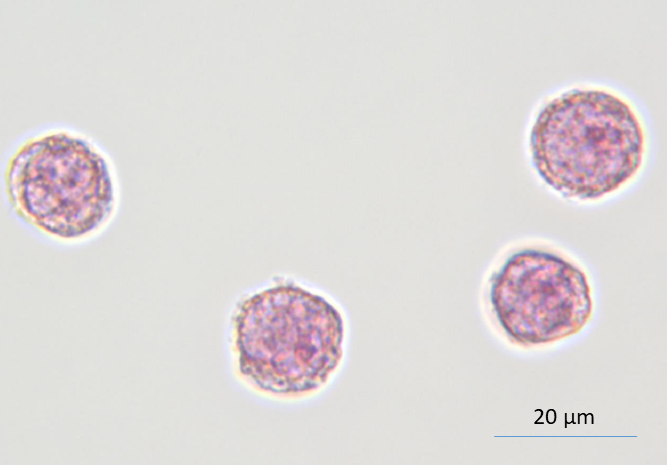} }}%
    \qquad
    \subfloat[\centering ]{{\includegraphics[width=0.45\textwidth]{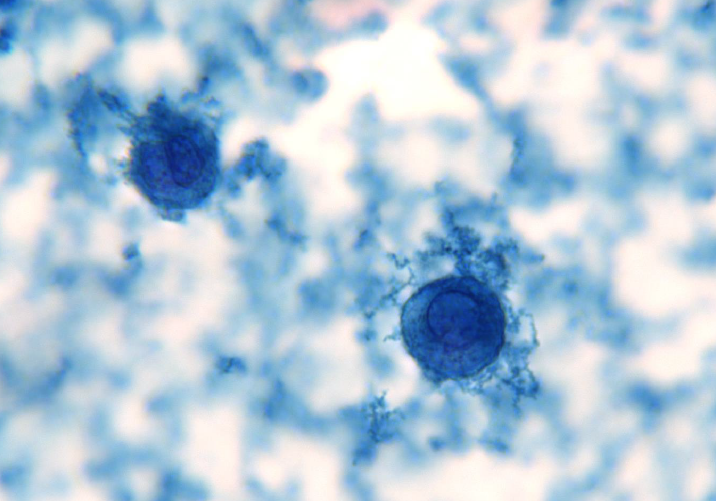} }}%
    \caption{THP-1 cells without (a) and with (b) addition of Perimag particles after 24 hours of incubation time. Iron in the sample is visualized by Prussian Blue staining. The particles are to a great extent taken up by the cells. Redundant particles outside the cells still form aggregates, being attached to the outer walls of the cells.}
    \label{Fig:PBS}
\end{figure}
\begin{figure}[h!]
  \centering
    \includegraphics[width=1\textwidth]{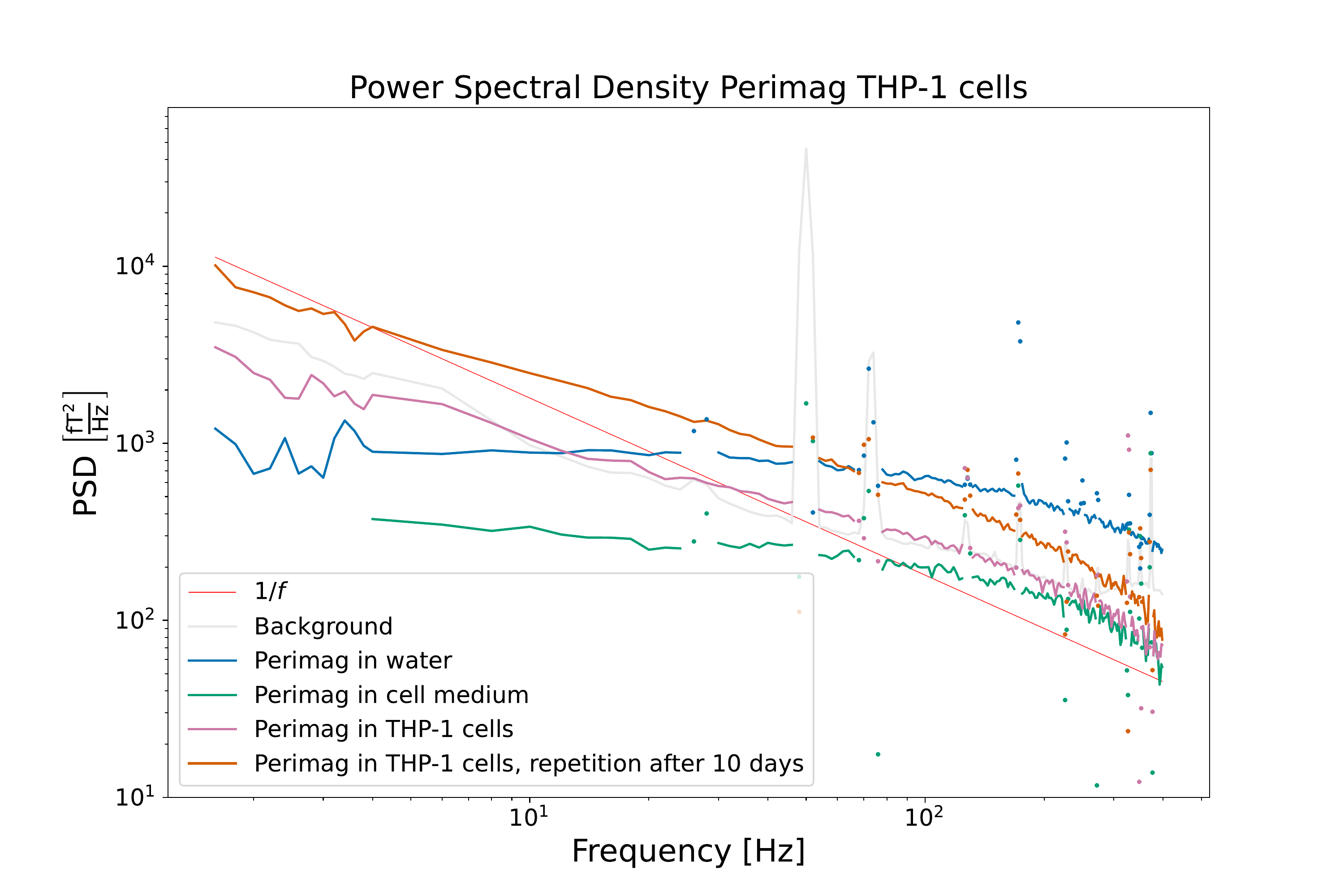}
    \caption{Power Spectral Densities of COOH coated Perimag particles before and after cellular uptake by THP-1 cells. The influence of the cell medium on the dynamics of the particles is very limited. Due to cluster formation and partial immobilization of the particles  after cellular uptake, the power in the lower frequency range increases. Full immobilization is however excluded, since the distinctive $1/f$ behaviour of Fig. \ref{Fig:ethanol_polymer} is not reached.}
    \label{Fig:cells}
\end{figure}

200 $\mu$L COOH coated Perimag particles with a concentration of c(Fe)=244.7 mmol/L were incubated with $2\cdot 10^7$ THP-1 cells in a 800 $\mu$l RPMI +1$\%$ FCS medium for 24 hours. Fig. \ref{Fig:PBS} showes the cells before and after the incubation (undiluted sample), where iron is visualized by Prussian blue. From these pictures, it is clear that the magnetic nanoparticles are taken up by the cells to a great extent. Moreover, MNP in the surrounding solution also form aggregates.

3 different samples have been measured in the tabletop TNM setup and are displayed in Fig. \ref{Fig:cells} with their respective colours:
\begin{enumerate}[label=\Alph*]
	\item Perimag particles in water suspension (blue)
	\item Perimag particles in the cell medium (green)
	\item Perimag particles after 24 hours of incubation time with THP-1 cells (pink).
\end{enumerate}
The PSD of the particles in the water suspension and the cell medium show only quantitative differences, which are due to the difference in concentration of the samples. The influence of the cell medium on the dynamics of the particles is - at least in the measured frequency range - very limited. After cellular uptake of the particles by the cells, a higher relative noise power in the lower frequency regime is measured, and the faster fluctuations are less present in the noise density. The broad distribution of cutoff frequencies clearly shifts towards lower values, which can be attributed to the formation of clusters and partial immobilization. Full immobilization can however be excluded, since the PSD does not fall off with $1/f$ as in Fig. \ref{Fig:ethanol_polymer} (a) and (b). A repetition measurement of Sample C was carried out after 10 days (orange curve). Apart from an increased SNR - which is related to the further cell sedimentation and a decreased average sample sensor distance - no qualitative differences were detectable. The magnetization dynamics of the particles in the well-aged cell sample shows no notable differences with that of the sample directly after the 24 hours incubation.

Two noise curves can be compared quantitatively, namely those of the particles suspended in the cell medium and those of the particles directly after the incubation time, because the MNP concentration and sample volume were similar. A continuous probing of the noise power at e.g. 10 Hz could quantify the cellular uptake during the incubation process.

\section{Conclusion and outlook}
The OPM based tabletop setup offers a flexible measurement unit to track changes in the thermal noise spectra of magnetic nanosystems. This flexibility provides the potential for use in processes and industrial application beyond biomedicine, such as 3d additive manufacturing. The power spectral densities of two commercially available MNP systems were compared with TNM measurements in a SQUID setup and a good agreement of the noise curves was found. These are the first thermal noise spectra of MNP ensembles measured with OPMs. Moreover, the proposed setup is particularly suited to monitor clustering and immobilization processes of the particles over time, due to the excellent performance of the OPMs in the lower frequency regime and the dependency of the TNM signal on the power of the individual fluctuators' volume. Three proof-of-concept experiments were performed to show their effect on the noise spectra of the particles. The immobilization of the particles induces a distinct $1/f$ dependency in the power spectral density, which results from the broad distribution of the N\'{e}el fluctuation times, as is visible from the gradual formation of UV polymers in an MNP sample. In contrast, the clustering of the particles when taken up by THP-1 cells slows down the Brownian fluctuations, due to the increased volume of the fluctuators, with a shift of the corresponding cutoff frequency towards lower frequencies as a result.

Presently, for a detailed MNP characterization by TNM, the SQUID setup remains preferred because of its broad bandwidth, which allows for the mapping of a broad range of MNP systems, and higher sensitivity, which facilitates the investigation of lower concentrated samples. However, a specifically for TNM designed OPM sensor could be more suited than the broadly applicable commercial magnetometers used here. By tuning the modulation frequency of the sensors, an optimum for the trade-off between sensitivity and bandwidth can be defined for each different MNP system. Further work will be dedicated to this.

\section*{Acknowledgments}
This work was supported by the German Research Foundation (DFG) through the Project ‘‘MagNoise: Establishing Thermal Noise
Magnetometry for Magnetic Nanoparticle Characterization’’ under Grant FKZ WI4230/3-1. J. L. was supported by the Fonds Wetenschappelijk Onderzoek (FWO-Vlaanderen) with senior postdoctoral research fellowship No. 12W7622N.

\bibliographystyle{unsrt}

\end{document}